%
%
%
%
%
%
%
%
%
%
\hoffset=0.0in
\voffset=0.0in
\hsize=6.5in
\vsize=8.9in
\normalbaselineskip=12pt
\normalbaselines
\topskip=\baselineskip
\parindent=15pt
%
%
%

\let\gm=\gamma
\let\dl=\delta

\let\om=\omega

\let\la=\langle
\let\ra=\rangle
\let\pa=\partial
\let\lf=\left
\let\rt=\right
\let\dt=\cdot
\let\del=\nabla
\let\dg=\dagger

\let\q=\widehat

\let\h=\hbar

\let\x=\times

\let\ty=\textstyle
\let\sy=\scriptstyle
\let\ssy=\scriptscriptstyle
\let\:=\>
\let\\=\cr
\let\emph=\e
\let\em=\it
\let\m=\hbox

\let\cl=\centerline

\def\e#1{{\it #1\/}}
\def\textbf#1{{\bf #1}}
\def\[{$$}
\def\]{\[}
\def\re#1#2{$$\matrix{#1\cr}\eqno{({\rm #2})}$$}

\def\tint{{\ty\int}}

\def\eqdf{\buildrel{\rm def}\over =}
\def\hf{{\sy {1 \over 2}}}
\def\qr{{\sy {1 \over 4}}}

\def\hfs{{\ssy {1 \over 2}}}
\def\qrs{{\ssy {1 \over 4}}}
\def\tqrs{{\ssy {3 \over 4}}}
\def\id{{\rm I}}

\def\qH{\q{H}}

\def\mathrm#1{{\rm #1}}

\def\mathcal#1{{\cal #1}}

\def\cE{{\cal E}}
\def\cH{{\cal H}}

\def\mbf{\fam\bffam\tenbf}
\def\bv#1{{\mbf #1}}

\def\vr{\bv{r}}

\def\vp{\bv{p}}

\def\vE{\bv{E}}
\def\vB{\bv{B}}
\def\vA{\bv{A}}

\def\vk{\bv{k}}

\def\vPsi{\bv{\Psi}}

\def\qvp{\q{\vp}}

\def\qvPsi{\q{\vPsi}}
\def\qq{\q{q}}
\def\qp{\q{p}}

\def\qpsi{\q{\psi}}
\def\qpsid{\qpsi\,{}^\dg}

\def\Schr{Schr\"{o}\-ding\-er}
\font\frtbf = cmbx12 scaled \magstep1
\font\twlbf = cmbx12
\font\ninbf = cmbx9
\font\svtrm = cmr17
\font\twlrm = cmr12
\font\ninrm = cmr9

\def\abstract#1{{\ninbf\cl{Abstract}}\medskip
\openup -0.1\baselineskip
{\ninrm\leftskip=2pc\rightskip=2pc #1\par}
\normalbaselines}
\def\sct#1{\vskip 1.33\baselineskip\noindent{\twlbf #1}\medskip}

\def\so{\raise 0.65ex \m{\sevenrm 1}}
\def\sk{\par\vskip 0.66\baselineskip}
{\svtrm
\cl{The Schr\"{o}dinger-equation presentation of}
\medskip
\cl{any oscillatory classical linear system}
\medskip
\cl{that is homogeneous and conservative}
}
\bigskip
{\twlrm
\cl{Steven Kenneth Kauffmann}
}
\cl{American Physical Society Senior Life Member}
\medskip
\cl{Unit 802, Reflection on the Sea}
\cl{120 Marine Parade}
\cl{Coolangatta QLD 4225}
\cl{Australia}
\cl{Tel: +61 4 0567 9058}
\smallskip
\cl{Email: SKKauffmann@gmail.com}
\bigskip\smallskip
\abstract{
The time-dependent Schr\"{o}dinger equation with time-independent Hamiltonian
matrix is a homogeneous linear oscillatory system in canonical form.  We
investigate whether any classical system that itself is linear, homogeneous,
oscillatory and conservative is guaranteed to linearly map into a
Schr\"{o}dinger equation.  Such oscillatory classical systems can be analyzed
into their normal modes, which are mutually independent, uncoupled simple
harmonic oscillators, and the equation of motion of such a system linearly maps
into a Schr\"{o}dinger equation whose Hamiltonian matrix is diagonal, with $h$
times the individual simple harmonic oscillator frequencies as its diagonal
entries.  Therefore if the coupling-strength matrix of such an oscillatory
system is presented in symmetric, positive-definite form, the Hamiltonian
matrix of the Schr\"{o}dinger equation it maps into is $\h$ times the square
root of that coupling-strength matrix.  We obtain a general expression for
mapping this type of oscillatory classical equation of motion into a
Schr\"{o}dinger equation, and apply it to the real-valued classical
Klein-Gordon equation and the source-free Maxwell equations, which results in
relativistic Hamiltonian operators that are strictly compatible with the
correspondence principle.  Once such an oscillatory classical system has been
mapped into a Schr\"{o}dinger equation, it is automatically in canonical form,
making second quantization of that Schr\"{o}dinger equation a technically
simple as well as a physically very interpretable way to quantize the original
classical system.
}

\sct{Introduction}
\noindent
A time-dependent \Schr\ equation, viewed as a real-valued equation of
motion that couples the real and imaginary parts of its wave vector,
is a homogeneous linear oscillatory canonical classical system (its
classical Hamiltonian function is the presentation in the appropriate
real canonical variables of the quantum expectation value of its Ham%
iltonian matrix).  Here we shall see that a \e{general} homogeneous
linear oscillatory conservative classical system's equation of motion
can always be linearly mapped into a \Schr\ equation, and that this
mapping is invertible if the classical system has no zero-frequency
normal modes.  When the oscillatory classical system's \e{coupling-%
strength matrix} is presented in \e{symmetric form} and is positive
definite (i.e., no zero-frequency normal modes), the corresponding
\Schr\ equation's \e{quantum Hamiltonian matrix} comes out to be $\h$
times the positive-definite \e{square root} of that classical coup%
ling-strength matrix.  The \Schr\ equation's complex-valued mapped
wave vector can be sensibly \e{normalized} such that the quantum sys%
tem's energy expectation value equals the oscillatory classical sys%
tem's energy function.  Moreover, that classical system can then be
immediately \e{quantized} by means of the very straightforward \e{sec%
ond quantization} of the \Schr\ equation that it maps into, which by
its \e{nature} is in \e{canonical form}.  This is not only technically
simple, it is as well automatically accompanied by a detailed physical
interpretation---e.g., one has a mathematical depiction of classical-%
wave/quantum-particle \e{complementarity} via the linear mapping of
the original classical \e{oscillatory} degrees of freedom (which have
Hermitian representation) into the second-quantized \Schr-equation
wave vector's \e{annihilation and creation} components (which have
non-Hermitian representation).  Mapping into a \Schr\ equation of the
real-valued classical scalar-field Klein-Gordon equation with mass pa%
rameter $m$ yields a complex-valued scalar wave function and the Ham%
iltonian operator $(|c\qvp|^2 + m^2c^4)^\hf$, which is in accord with
the correspondence-principle prescription for a relativistic free par%
ticle of mass $m$~[1].  Such mapping of the classical source-free Max%
well equations yields a complex-valued transverse-vector wave function
and the Hamiltonian operator $c|\qvp|$, which is relativistically ap%
propriate to the massless free photon~[2].

Oscillatory classical linear systems which are homogeneous and conser%
vative are described by second-order equations of motion that have the
form,
\re{
    \ddot q + Kq = 0,
}{1a}
where $K$ is a nonvanishing real-valued matrix, all of whose eigen%
values are real and nonnegative, and whose real-valued eigenvectors
\e{completely span} the real-valued vector space on which $K$ natural%
ly operates.  Note that the use of the terms ``vector'' and ``matrix''
in this article is \e{not} intended to exclude vectors that have a
continuum of components (e.g., functions) or matrices that have a con%
tinuum of entries (e.g., operators on function spaces).  However, in
the interest of cutting down on notational clutter, all of the didac%
tic \e{generic formulas} that are presented in this article which in%
volve vector components or matrix entries display \e{only} the case
that these are discrete---that is notwithstanding the fact that the
interesting \e{examples} which are discussed in the last part of this
article all have continuum character.

We note that \e{first-order} classical equations of motion which have
the simple form,
\re{
    \dot s = Ws,
}{1b}
\e{also} describe homogeneous linear oscillatory conservative classi%
cal systems when $W$ is a nonvanishing real-valued matrix that has ex%
clusively \e{imaginary} eigenvalues whose associated complex-valued
eigenvectors \e{completely span} the \e{extended} complex-valued vec%
tor space on which the real-valued $W$ can operate. This is so because
Eq.~(1b) implies that,
\re{
    \ddot s - W^2s = 0,
}{1c}
and the matrix $-W^2$ can be shown to conform to all the requirements
stipulated for the matrix $K$ below Eq.~(1a).  To see this, note that
if $s_{\om}$ is \e{any} complex-valued eigenvector of $W$, with $i\om$
its corresponding imaginary eigenvalue, where $\om$ is a real number,
then \e{because} $W$ is \e{real-valued}, the particular \e{complex-%
conjugated} vector $s_{\om}^\ast$ is \e{as well} an eigenvector of
$W$, but with eigenvalue $-i\om$.  Therefore the \e{real-valued} vec%
tor $s_{\om} + s_{\om}^\ast$ is an eigenvector of the real-valued ma%
trix $-W^2$ with the real, nonnegative eigenvalue $\om^2$.  In addi%
tion, since the complex-valued eigenvectors of $W$ of the form
$s_{\om}$ are assumed to completely span the extended complex-valued
vector space on which $W$ can operate, it is apparent that the real-%
valued eigenvectors of $-W^2$ that have the form $s_{\om} +
s_{\om}^\ast$ completely span the \e{real-valued} vector space on
which the real-valued matrix $-W^2$ naturally operates---and of course
the \e{eigenvalues} $\om^2$ of $-W^2$ associated to each member of
this \e{complete set} of its real-valued eigenvectors are themselves
real-valued and \e{nonnegative}.  Therefore the nonvanishing real-val%
ued matrix $-W^2$ of Eq.~(1c) possesses \e{all} of the properties that
are required of the real-valued matrix $K$ of Eq.~(1a).

It is further to be noted at this point that if the nonvanishing real%
-valued matrix $W$ is \e{antisymmetric}, then it \e{automatically}
fulfills the remaining requirements that are stipulated below Eq.~%
(1b), and, \e{in addition}, a linear mapping of Eq.~(1b) into a \Schr\
equation is immediately manifest.  This is so because if $W$ is real-%
valued and \e{antisymmetric}, then $iW$ is \e{Hermitian} on the exten%
ded complex-valued vector space on which $W$ can operate.  By virtue
of its Hermitian property, $iW$ necessarily possesses a \e{complete
set} of complex-valued eigenvectors, for each of which it has a cor%
responding \e{real} eigenvalue.  Those real eigenvalues of $iW$ cor%
respond, of course, to \e{imaginary} eigenvalues of $W$ with the
\e{same} corresponding eigenvectors, and that set of eigenvectors of
course completely spans the extended complex-valued vector space on
which $W$ can operate.  In addition, if we multiply both sides of Eq.%
~(1b) by the factor $i\h$, it becomes a \Schr\ equation with the Herm%
itian Hamiltonian matrix $i\h W$.

In the next section we shall show that classical equations of motion
given by Eq.~(1a), with the restrictions on the matrix $K$ that are
stipulated below Eq.~(1a), can always be linearly mapped into \Schr\
equations---consequently the same is true for classical equations of
motion given by Eq.~(1b) with the restrictions on the matrix $W$ that
are stipulated below Eq.~(1b)).  This task will be greatly facilitated
by the fact that the oscillatory classical Eq.~(1a) can be analyzed
into its \e{normal modes}, which, of course, behave as \e{mutually in%
dependent simple harmonic oscillators}.  It turns out that a classical
simple harmonic oscillator equation of motion which has the natural
angular frequency $\om$ can be linearly mapped into a \Schr\ equation
for an ultra-basic \e{single-state} quantum system whose \e{one-by-%
one} Hamiltonian ``matrix'' is either the real number $\h\om$ or the
real number $-\h\om$.  The classical equation of motion for a \e{col%
lection} of such \e{mutually independent} simple harmonic oscillators
(i.e., an oscillatory classical system that has been \e{analyzed} into
its normal modes) correspondingly linearly maps into a \Schr\ equation
whose Hamiltonian matrix is \e{diagonal}, with its diagonal entries
corresponding in one-to-one fashion to the angular frequencies of the
independent simple harmonic oscillators which comprise that particular
collection: each such Hamiltonian-matrix diagonal entry is a unique
one of those angular frequencies times one of the two allowed factors
$\pm\h$.

We now turn to the technical details of the analysis of Eq.~(1a) into
its normal modes, and the subsequent linear mapping of such collec%
tions of independent simple harmonic oscillators into \Schr\ equa%
tions.

\sct{Analysis into normal modes and their mapping into \Schr\ equations}
\noindent
The real-valued eigenvectors $q_j$ of $K$ in Eq.~(1a) completely span
the real-valued vector space on which $K$ naturally operates, and each
$q_j$ corresponds to a nonnegative eigenvalue $\om_j^2$, where we take
$\om_j$ to be real and nonnegative.  Therefore the $q_j$ satisfy ei%
genvalue equations of the form,
\re{
    Kq_j = \om_j^2q_j
}{2a}
It turns out that we can use these eigenvectors $q_j$ to construct a
matrix $S$ which is invertible, and for which the composite matrix
$S^{-1}KS$ is in \e{diagonal form}, with all of its nondiagonal en%
tries being equal to zero, while its diagonal entries embrace all the
eigenvalues $\om_j^2$ of $K$.  \e{Because} of this \e{diagonal form}
of the matrix $S^{-1}KS$, it will be the case that \e{each} of the
\e{components} $(S^{-1}q)_j$ of the \e{transformation} $S^{-1}q$ of
the dynamical vector $q$ of Eq.~(1a) satisfies an \e{independent} sim%
ple harmonic oscillator equation whose natural angular frequency
$\om_j$ is the nonnegative square root of one of the eigenvalues
$\om_j^2$ of the matrix $K$.  In short, the \e{components} of the
\e{transformed vector} $S^{-1}q$ are the \e{normal modes} of Eq.~(1a).

We shall now construct the matrix $S$ by filling its columns with the
components of a set of linearly independent $q_j$, where that set is
sufficiently large to completely span the real-valued vector space on
which K naturally operates,
\re{
    S_{ij}\eqdf(q_j)_i.
}{2b}
Because the columns of the matrix $S$ are linearly independent and
completely span the real-valued vector space on which $S$ (and $K$)
naturally operate, the matrix $S$ will have an \e{inverse} $S^{-1}$. 
In addition, because of the eigenvalue equations given by Eq.~(2a) and
the definition of $S$ given by Eq.~(2b), it is readily verified that,
\re{
    (KS)_{kj} = K_{kl}(q_j)_l = (Kq_j)_k = \om_j^2 S_{kj}.
}{2c}
This result permits us to verify that $S^{-1}KS$ is precisely the dia%
gonal form of the matrix $K$ mentioned below Eq.~(2a),
\re{
    (S^{-1}KS)_{mj} = (S^{-1})_{mk}(KS)_{kj} =
    \om_j^2(S^{-1})_{mk}S_{kj} = \om_j^2(S^{-1}S)_{mj} = \om_j^2\dl_{mj}.
}{2d}
It is convenient to denote this diagonal form of $K$ as $K_S$,
\re{
    K_S\eqdf S^{-1}KS,
}{3a}
If we now multiply Eq.~(1a) through by the matrix $S^{-1}$ and further
define,
\re{
    q_S\eqdf S^{-1}q,
}{3b}
we obtain from Eq.~(1a) that,
\re{
    \ddot{q}{}_S + K_Sq_S = 0,
}{3c}
which, from Eqs.~(3a) and (2d), reads when written in component form,
\re{
    d^2(q_S)_j/dt^2 + \om_j^2(q_S)_j = 0,
}{3d}
which is a set of mutually independent simple harmonic oscillator
equations whose natural angular frequencies $\om_j$ are given by
the \e{nonnegative square roots} of the nonnegative \e{eigenvalues}
$\om_j^2$ of the matrix K.  From Eq.~(3d) we see that the normal-mode
simple harmonic oscillator variables are the \e{components} of the
vector $q_S$.

We wish at this point to further linearly map the set of mutually in%
dependent simple harmonic oscillator equations encompassed by Eq.~(3c)
into a \Schr\ equation.  Notwithstanding that they have the \e{same
form}, Eqs.~(3c) and (1a) \e{crucially differ} in that $K_S$ in Eq.~%
(3c) is \e{known to be diagonal} (with real nonnegative entries on the
principal diagonal and uniformly zero entries elsewhere), whereas $K$
in Eq.~(1a) is \e{not} guaranteed to be diagonal.  When we now attempt
to pass to a \Schr\ equation, we obviously do \e{not} wish to \e{undo}
the \e{simplicity} that having \e{only diagonal matrices present} con%
fers on an equation of motion.  Therefore we now make it a rigid rule
that \e{any} attempted further linear mapping of Eq.~(3c) into (hope%
fully) a \Schr\ equation \e{may only be attempted with diagonal matri%
ces}.  This affords an \e{immediate benefit}: diagonal matrices \e{all
mutually commute}.

Now a \Schr\ equation has the form,
\re{
    i\h\dot\psi = H\psi,
}{4a}
and, of course, our \e{cardinal rule} stated above requires that the
Hermitian matrix $H$ be \e{diagonal}.

Eq.~(3c) is second-order in time, whereas the \Schr\ Eq.~(4a) is
first-order in time.  To \e{reconcile} this difference in order, it is
\e{necessary} to take $\psi$ to be a linear mapping of ${\dot q}_S$,
and possibly of $q_S$ itself as well.  Therefore we now make the
\e{ansatz},
\re{
    \psi = iN(Wq_S + {\dot q}_S),
}{4b}
where the matrices $N$ and $W$ are of course \e{both} assumed to be
\e{diagonal}.  We make the further assumption that the matrix $N$ is
invertible (i.e., has \e{no} vanishing entries on its principal dia%
gonal), which implies that it simply \e{factors out} of the linear,
homogeneous \Schr\ Eq.~(4a), and therefore is \e{not} determined by
it.  From Eq.~(3c), we know that ${\ddot q}_S = -K_Sq_S$.  Therefore
putting the \e{ansatz} of Eq.~(4b) into the \Schr\ Eq.~(4a) results
in,
\re{
    i\h W{\dot q}_S - i\h K_Sq_S = HWq_S + H{\dot q}_S,
}{4c}
which yields the two equations,
\re{
    i\h W = H, \quad -i\h K_S = HW,
}{4d}
that have the solutions,
\re{
    H = \h (K_S)^\hfs, \quad W = -i (K_S)^\hfs,
}{4e}
which are \e{consistent} with our \e{assumption} that $H$ and $W$ are
\e{diagonal matrices}, and \e{also} imply that $H$ is Hermitian.  Put%
ting the results of Eq.~(4e) into Eq.~(4b) together with the defini%
tion of $K_S$ given by Eq.~(3a) and that of $q_S$ given by Eq.~(3b)
yields the desired linear mapping of $q$ and $\dot q$ of Eq.~(1a) into
the \Schr\ equation wave vector $\psi$, and also yields the associated
Hamiltonian matrix H of that \Schr\ equation,
\re{
    \psi = N((S^{-1}KS)^\hfs S^{-1}q + iS^{-1}\dot{q}),\quad
       H = \h (S^{-1}KS)^\hfs.
}{4f}
From Eq.~(4f), bearing in mind that both $N$ and $(S^{-1}KS)^\hfs$ are
mutually commuting \e{diagonal} matrices and $N$ is invertible, it can
readily be shown that the \Schr\ Eq.~(4a) for $\psi$ \e{follows} from
the underlying \e{classical} Eq.~(1a) for $q$.

We as well note from Eq.~(4f) that if all the eigenvalues of K are
\e{positive}, i.e., the classical system is \e{purely} oscillatory,
then the diagonalized matrix $S^{-1}KS$ is \e{invertible}, as is the
diagonal matrix $(S^{-1}KS)^\hfs$, and therefore the linear mapping
between $q$ and $\psi$ is \e{also} invertible.

An interesting mathematical point is that since the diagonal entries
of $S^{-1}KS$ are all real and nonnegative (they are the the eigenval%
ues of $K$), $(S^{-1}KS)^\hfs$ is certainly \e{defined} as a diagonal
matrix, but \e{multiply} so, i.e., the \e{signs} of the nonzero diag%
onal entries of $(S^{-1}KS)^\hfs$ \e{can be chosen at will}.  So from
a strictly mathematical point of view, Eq.~(4f) specifies a whole set
of distinct linear mappings of the classical $q$ into \Schr\ wave vec%
tors $\psi$, with \e{equally distinct} Hamiltonian matrices $H = \h
(S^{-1}KS)^\hfs$ to accompany each distinct linear mapping.

Although the \Schr\ Eq.~(4a) \e{does not determine} the invertible di%
agonal ``normalization'' matrix $N$ of our \Schr\ wave vector $\psi$
of Eq.~(4f), we can ask if there is an \e{additional physically sensi%
ble requirement} which impinges on the value of that ``normalization''
diagonal matrix $N$.

Now the behavior of quantum expectation values frequently closely par%
allels that of their classical counterparts, as Ehrenfest's Theorem
attests, and that is particularly the case for simple linear systems.
Specifically, the expectation value of the Hamiltonian matrix $H$,
namely $\psi^\ast H\psi$, is a \e{real-valued} function of $\psi$ and
$\psi^\ast$ with the dimension of \e{energy} which is \e{conserved}
because the time evolution of $\psi$ is governed by the \Schr\ Eq.~%
(4a) and the Hamiltonian matrix is Hermitian---this \e{conservation}
of $\psi^\ast H\psi$ can be explicitly verified.  The clear classical
analog of $\psi^\ast H\psi$ is therefore, of course, the \e{classical
conserved energy} that is associated with the Eq.~(3c) classical equa%
tion of motion.  That classical conserved energy is the nonnegative
entity,
\re{
    \cE_{K_S}(q_S, \dot q_S)\eqdf (\dot q_S\dot q_S + q_SK_Sq_S)/(2\gm^2),
}{5a}
where the dimension and magnitude of the real positive number $\gm$
depends on the dimension and normalization of $q_S$---note that
$\cE_{K_S}(q_S, \dot q_S)$ is \e{required} to have the dimension of
\e{energy}.  That $\cE_{K_S}(q_S, \dot q_S)$ is \e{conserved}, i.e.,
that its time derivative \e{vanishes}, follows directly from  Eq.~(3c)
itself and the fact that $K_S$ is diagonal.

Therefore it is \e{completely sensible physically} to attempt to de%
termine $N$ by \e{additionally} imposing the utterly natural require%
ment that,
\re{
    \psi^\ast H\psi = \cE_{K_S}(q_S, \dot q_S) =
    (\dot q_S\dot q_S + q_SK_Sq_S)/(2\gm^2),
}{5b}
whenever this is possible---we shall see that Eq.~(5b) requires the
real nonnegative diagonal matrix $K_S$ to be positive definite, i.e.,
the classical system must be \e{purely} oscillatory.  Furthermore, the
strictly \e{nonnegative character} of the classical energy
$\cE_{K_S}(q_S, \dot q_S)$ now \e{precludes the possibility} that the
diagonal Hamiltonian matrix $H = \h (S^{-1}KS)^\hfs$ can have \e{any%
thing other than nonnegative entries}.  \e{Unlike} the \Schr\ Eq.%
~(4a), Eq.~(5b) is, of course, \e{neither} linear nor homogeneous in
$\psi$.  We now reexpress Eq.~(4f) in the more compact form,
\re{
    \psi = N((K_S)^\hfs q_S + i\dot q_S),\quad
       H = \h(K_S)^\hfs,
}{5c}
and substitute the right-hand sides of both the first and second e%
qualities of Eq.~(5c) into the left hand side of Eq.~(5b).  For the
left and right hand sides of Eq.~(5b) to then be able to be equal, the
following equation involving the diagonal matrices $N^\ast$, $N$ and
$(K_S)^\hfs$ must be satisfied,
\re{
    N^\ast N(K_S)^\hfs = \id/(2\h\gm^2),
}{5d}
where $\id$ is the identity matrix.  Of course this is \e{not possi%
ble} if $(K_S)^\hfs$ has \e{any} vanishing or negative diagonal en%
tries.  If $(K_S)^\hfs$ indeed has \e{only positive entries}, which
can only be the case if the classical system is \e{purely} oscillator%
y, then the \e{simplest} solution for the diagonal matrix $N$ is one
with only real-valued diagonal entries, namely,
\re{
    N = (K_S)^{-\qrs}/(2\gm^2\h)^\hfs.
}{5e}
Putting this determination of $N$ into Eq.~(5c) results in the proper%
ly normalized \Schr\ wave vector,
\re{
    \psi = ((K_S)^\qrs q_S + i(K_S)^{-\qrs}\dot q_S)/(2\gm^2\h)^\hfs,\quad
     H = \h (K_S)^\hfs,
}{5f}
which in the more explicit notation used in Eq.~(4f) reads,
\re{
    \psi = ((S^{-1}KS)^\qrs S^{-1}q +
    i(S^{-1}KS)^{-\qrs}S^{-1}\dot q)/(2\gm^2\h)^\hfs,\quad
     H = \h(S^{-1}KS)^\hfs,
}{5g}
where the diagonal matrices $S^{-1}KS$ and $(S^{-1}KS)^\hfs$ now both
need to be \e{positive definite}, and the real positive constant $\gm$
comes from the classical energy function $\cE_{S^{-1}KS}(S^{-1}q,
S^{-1}\dot q)$ of Eq.~(5a) that is appropriate to the \e{purely} os%
cillatory classical equation of motion system of Eq.~(3c),
\re{
    \cE_{S^{-1}KS}(S^{-1}q, S^{-1}\dot q) = 
     ((S^{-1}\dot q)(S^{-1}\dot q) + (S^{-1}q)(S^{-1}KS)(S^{-1}q))/(2\gm^2),
}{5h}
Because the diagonal matrix $(S^{-1}KS)^\hfs$ is positive definite,
the linear mapping of $q$ and $\dot q$ into $\psi$ given in Eq.~(5g)
is \e{invertible},
\re{
    q =   ((\gm^2\h)/2)^\hfs S(S^{-1}KS)^{-\qrs}(\psi + \psi^\ast),\quad
  \dot q = -i((\gm^2\h)/2)^\hfs S(S^{-1}KS)^\qrs(\psi - \psi^\ast).
}{5i}

While the Eq.~(5g) route to the desired invertible linear mapping of
Eq.~(1a) into \Schr\ Eq.~(4a) is of great generality in principle, in
practice it suffers from the need to explicitly know all the eigenvec%
tors of $K$ in order to be able to \e{construct} $S$, and, in addi%
tion, from the need to explicitly \e{invert} $S$.

We see from Eq.~(5g) that one of the consequences of having the matrix
$S$ and its inverse $S^{-1}$ in hand is that the \Schr\ equation's
Hamiltonian matrix $H = \h(S^{-1}KS)^\hfs$ is presented to us in
\e{already diagonal} form.  It is certainly \e{not essential} that
that be the case.  In the next section we therefore simply expunge $S$
and its inverse from Eq.~(5g), which of course will work if $K$ is
\e{diagonal}.  However it quickly becomes clear that the resulting
expression \e{still} works when $K$ is merely \e{symmetric}.

\sct{\Schr-equation presentation of symmetrically coupled oscillatory systems}
\noindent
The result of expunging $S$ and $S^{-1}$ from Eq.~(5g) is,
\re{
    \psi = (K^\qrs q + iK^{-\qrs}\dot q)/(2\gm^2\h)^\hfs,\quad
     H = \h K^\hfs,
}{6a}
and if $K$ is a real-valued \e{symmetric} positive-definite matrix,
all the expressions in it \e{still} make sense: in those circumstances
$H = \h K^\hfs$ is well defined as a real-valued \e{symmetric} posi%
tive-definite matrix \e{itself}.  Therefore $H$ is \e{Hermitian}, as
required, and $K^\qrs$ and $K^{-\qrs}$ are well-defined as real-valued
symmetric invertible matrices.  Furthermore, it is straightforwardly
verified that in \e{consequence} of the basic oscillatory classical e%
quation of motion of Eq.~(1a), the wave vector $\psi$ of Eq.~(6a)
\e{satisfies} the \Schr\ Eq.~(4a) with the Hamiltonian matrix $H =
\h K^\hfs$ given by Eq.~(6a).  In addition, when $K$ is a real-valued
\e{symmetric} positive-definite matrix, Eq.~(6a) yields,
\re{
    \psi^\ast H\psi = (\dot q\dot q + qKq)/(2\gm^2),
}{6b}
and if $\gm$ has been appropriately selected such that,
\re{
    \cE_K(q, \dot q)\eqdf(\dot q\dot q + qKq)/(2\gm^2),
}{6c}
has the dimension of \e{energy}, then it is clear that,
\re{ 
    L_K(q, \dot q)\eqdf(\dot q\dot q - qKq)/(2\gm^2),
}{6d}
\e{also} has the dimension of energy.  Moreover, it is easily verified
that the Euler-Lagrange equation which \e{follows} from the Lagrangian
$L_K(q, \dot q)$ of Eq.~(6d) is \e{precisely} the Eq.~(1a) classical
equation of motion.  Now the \e{conserved energy} of any classical
system that \e{has} a Lagrangian $L$ is well-known to be \e{uniquely
given} by $(\dot q\del_{\dot q}L - L)$, which, for the particular
Eq.~(1a) case that $L$ is given by $L_K(q, \dot q)$ of Eq.~(6d), is
straightforwardly verified to be $\cE_K(q, \dot q)$, as defined by
Eq.~(6c).  Therefore, Eqs.~(6b) and (6c) show that when $K$ is real-%
valued, \e{symmetric} and positive definite, then the expectation val%
ue of the Hamiltonian matrix which \e{follows} from Eq.~(6a) is \e{e%
qual} to the conserved energy of the \e{classical} system of Eq.~(1a),
as required.

Finally, when $K$ is real-valued, symmetric and positive definite, the
\e{inverse} of the Eq.~(6a) linear mapping of $q$ and $\dot q$ into
$\psi$ is readily calculated to be,
\re{
    q =   ((\gm^2\h)/2)^\hfs K^{-\qrs}(\psi + \psi^\ast),\quad
  \dot q = -i((\gm^2\h)/2)^\hfs K^\qrs(\psi - \psi^\ast),
}{6e}
which is, as expected, the result of expunging $S$ and $S^{-1}$ from
Eq.~(5i).

What if the matrix $K$ of the classical Eq.~(1a) is nonsymmetric?  We
then \e{first} need to find a real-valued invertible matrix $S$ such
that the \e{similarity-transformed} $K_S\eqdf S^{-1}KS$ \e{is} symmet%
ric.  Eq.~(6a) is \e{extended} to cover this situation by replacing
$K$ by $K_S$ and $q$ by $q_S\eqdf S^{-1}q$, \e{precisely} as in Eq.~%
(5f), \e{except} that \e{now} $K_S$ is \e{merely symmetric and posi%
tive definite}, not necessarily diagonal.

In the next section, we use the machinery of Eq.~(5f) and its associ%
ated Eq.~(3c) similarity-transformed version of the Eq.~(1a) classical
equation of motion (albeit always bearing in mind that $K_S$ is
\e{merely symmetric and positive definite}, not diagonal) to show that
the real and imaginary parts of $\psi$ times the factor $(2\h)^\hfs$
obey a simple first-order coupled equation of motion which can imme%
diately be Hamiltonized and then quantized.  This \e{second quantiza%
tion} of an oscillatory classical system's linear mapping into a
\Schr\ equation is a very easy route to that underlying system's quan%
tization, and one which as well automatically yields considerable
physical insight.

\sct{Hamiltonization and quantization of the \Schr-equation presentation}
\noindent
Taking the real-valued similarity-transformed $K_S$ in both Eqs.~(3c)
and (5f) to now be, as discussed above, \e{merely symmetric and posi%
tive definite} rather than necessarily diagonal, we note that the real
and imaginary parts of the wave vector $\psi$ of Eq.~(5f), each multi%
plied (for later convenience) by the factor $(2\h)^\hfs$, are given
by,
\re{
    q_c\eqdf(\h/2)^\hfs(\psi + \psi^\ast) = (K_S)^\qrs q_S/\gm,
    \qquad
    p_c\eqdf-i(\h/2)^\hfs(\psi - \psi^\ast) = (K_S)^{-\qrs}\dot q_S/\gm,
}{7a}
which are readily seen, as a consequence of Eq.~(3c), which is a simi%
larity-transformed version of the underlying Eq.~(1a) classical equa%
tion of motion, to satisfy the simple first-order coupled antisymme%
trical equation of motion,
\re{
    \dot q_c = (K_S)^\hfs p_c, \qquad \dot p_c = -(K_S)^\hfs q_c.
}{7b}
With a little effort, it can also be verified that the Eq.~(7b) system
\e{implies} the \Schr\ Eq.~(4a) with $H = \h(K_S)^\hfs$.  Moreover,
for $H = \h(K_S)^\hfs$, where $(K_S)^\hfs$ is real-valued and symmet%
ric, the two equalities of the Eq.~(7b) system \e{follow from} simply
the real and imaginary parts of the \Schr\ Eq.~(4a).  In other words,
for the situation that we are concerned with here, namely that $H =
\h(K_S)^\hfs$, where $(K_S)^\hfs$ is real-valued, symmetric and posi%
tive definite, the real-valued coupled antisymmetrical system of Eq.~%
(7b) is \e{completely equivalent} to the complex valued \Schr\ Eq.~%
(4a).

In \e{addition}, Eq.~(7b) \e{also} follows from a simple bilinear
\e{classical Hamiltonian}, namely,
\re{
    \cH_{K_S}(q_c, p_c) = (q_c(K_S)^\hfs q_c + p_c(K_S)^\hfs p_c)/2,
}{7c}
via the \e{classical canonical} Hamiltonian equations of motion, i.e.,
\re{
    \dot q_c =  \del_{p_c}\cH_{K_S}(q_c, p_c),\quad
    \dot p_c = -\del_{q_c}\cH_{K_S}(q_c, p_c),
}{7d}
and the fact that $(K_S)^\hfs$ is a real \e{symmetric} matrix.

By putting the definition of $(q_c, p_c)$ given in Eq.~(7a) into Eq.~%
(7c) we can reexpress our system's \e{classical Hamiltonian} in terms
of its \Schr-equation presentation wave vector $\psi$ and complex con%
jugate $\psi^\ast$,
\re{
    \cH_{K_S}(q_c, p_c) = (\psi^\ast H\psi + \psi H\psi^\ast)/2
                        =  \psi^\ast H\psi,
}{7e}
where the last equality in Eq.~(7e) follows from the fact that $H = 
\h (K_S)^\hfs$ is a real, \e{symmetric} matrix.  It is pleasing to
once again see the quantum expectation value of the Hamiltonian matrix
$H$ come out to be \e{equal} to the \Schr-equation presentation's
\e{classical} energy, i.e., to its \e{classical Hamiltonian}.

Since Eqs.~(7c) and (7d) assure us that the classical equations of mo%
tion of Eq.~(7b) obeyed by $(q_c, p_c)$ are \e{indeed} presented in
\e{canonical Hamiltonian form}, we can now safely \e{quantize} this
classical system by imposing first Dirac's canonical commutation rules
on the \e{components} of $(q_c, p_c)$, and next Heisenberg's equations
of motion on the now quantized $(\qq_c, \qp_c)$.  Dirac's canonical
commutation rules \e{promote} the \e{components} of $(q_c, p_c)$ into
noncommuting \e{Hermitian operators} which obey the commutation rela%
tions,
\re{
    [(\qq_c)_i, (\qq_c)_j] = [(\qp_c)_i, (\qp_c)_j] = 0,\quad
    [(\qq_c)_i, (\qp_c)_j] = i\h\dl_{ij}.
}{8a}

The Eq.~(8a) commutation relations, in turn, \e{imply} that the compo%
nents of the \e{non-Hermitian} quantized wave vector $\qpsi = (\qq_c +
i\qp_c)/(2\h)^\hfs$ satisfy, \e{in conjunction with} the components of
this quantized wave vector's \e{Hermitian conjugate} $\qpsid = (\qq_c
- i\qp_c)/(2\h)^\hfs$, the following commutation relations,
\re{
    [\qpsi_i, \qpsi_j] = [\qpsi_i\,{}^\dg, \qpsi_j\,{}^\dg] = 0,\quad
    [\qpsi_i, \qpsi_j\,{}^\dg] = \dl_{ij}.
}{8b}
These Eq.~(8b) commutation relations are the \e{fundamental} ones for
the components of our \Schr-equation presentation \e{quantized} wave
vector, and they give those quantized wave-vector components and their
Hermitian conjugates, respectively, their well-known interpretation as
\e{annihilation and creation operators}, which is so often crucial to
physical understanding.  They are \e{as well} the key to constructing
\e{physically useful} orthogonal basis sets for the \e{second-quan%
tized} Hilbert space that is the result of the  imposition of Dirac's
canonical commutation rules on the components of the dynamical-varia%
ble vector $(q_c, p_c)$.

The Hamiltonian \e{operator} for this quantized (i.e., \e{second}
quantized) \Schr-equation presented system is obtained by substituting
the \e{quantized} dynamical-variable vector $(\qq_c, \qp_c)$ into the
system's \e{classical} Hamiltonian of Eq.~(7c), namely by writing
down,
\re{
    \cH_{K_S}(\qq_c, \qp_c) = (\qq_c(K_S)^\hfs\qq_c + \qp_c(K_S)^\hfs\qp_c)/2,
}{8c}
which could have ambiguities due to \e{operator-ordering} issues, but
it is apparent that \e{those do not arise} in this case.  Noting that
the quantized dynamical-variable vector $(\qq_c, \qp_c)$ is given in
terms of the quantized wave vector $\qpsi$ and its Hermitian conjugate
$\qpsid$ by the quantized analog of the two definitions in Eq.~(7a),
namely $\qq_c = (\h/2)^\hfs(\qpsi + \qpsid)$ and $\qp_c = -i(\h/2)^%
\hfs(\qpsi - \qpsid)$, we reexpress the \e{uniquely defined} second-%
quantized Hamiltonian operator $\cH_{K_S}(\qq_c, \qp_c)$ of Eq.~(8c)
in terms of the \e{quantized wave vector} $\qpsi$ and its Hermitian
conjugate $\qpsid$,
\re{
    \cH_{K_S}(\qq_c, \qp_c) = (\qpsid H\qpsi + \qpsi H\qpsid)/2,
}{8d}
where $H = \h(K_S)^\hfs$, a real symmetric positive definite matrix.

If we now apply Heisenberg's equation of motion and the commutation
rules for the components of the quantized $\qpsi$ and $\qpsid$ that
are given by Eq.~(8b) to the second-quantized Hamiltonian operator
written in the form given by Eq.~(8d), we can calculate the time deri%
vative of any component of the \Schr-equation presentation \e{quan%
tized} wave vector $\qpsi$,
\re{
    d\qpsi_i/dt = (-i/\h)[\qpsi_i, (\qpsid H\qpsi + \qpsi H\qpsid)/2]
                =(-i/\h)((H\qpsi)_i + (\qpsi H)_i)/2 = (-i/\h)(H\qpsi)_i,
}{8e}
where the last step reflects the real \e{symmetric} character of the
Hamiltonian matrix $H = \h (K_S)^\hfs$.  Thus we have shown that,
\re{
   i\h d\qpsi/dt = H\qpsi,
}{8f}
i.e., the \Schr\ Eq.~(4a) which the \Schr-equation presentation wave
vector $\psi$ satisfies is \e{also} satisfied by that wave vector's
operator \e{quantization} $\qpsi$, which itself is, of course, a vec%
tor of the \e{annihilation operators} of the complete set of quantum
states which the \e{components} of the wave vector $\psi$ individually
describe.

We next turn to the \Schr-equation presentations of specifically the
classical Klein-Gordon equation and the source-free Maxwell equations.

\sct{The spinless quantum free particle from the classical Klein-Gordon
equation}
\noindent
The classical Klein-Gordon equation for the real-valued scalar field
$\phi$ differs from the classical wave equation by a simple mass
term~[3, 1],
\re{
\ddot\phi + (-c^2\del^2 + \om^2)\phi = 0,
}{9a}
where $\om = ((mc^2)/\h)$.  Eq.~(9a) has the form of Eq.~(1a) with,
\re{
    K = -c^2\del^2 + \om^2,
}{9b}
which, on the space of real-valued scalar fields, is a real-valued,
symmetric, positive-definite operator with the dimension of frequency
squared.  Therefore, starting with Eq.~(6a) above and going right
through to Eq.~(8f), we have results that can all be transcribed for
the real-valued classical Klein-Gordon equation.  We need to bear in
mind that during this exercise $K$ is specifically defined by Eq.~(9b)
and that the real-valued classical dynamical vector $q$ is defined by
the real-valued $\phi$, which, as a real-valued vector, of course has
a three-dimensional \e{continuous} index instead of a discrete one.
In such a case the \e{summation} that defines index contraction is
willy-nilly supplanted by three-dimensional \e{integration}, which
compels some systematic technical changes in the formalism, for exam%
ple in the dimension of the variables that one deals with (summation
is over dimensionless indices, integration here involves the three
space dimensions) and in the fact that Kronecker deltas give way to
three-dimensional delta functions.  That notwithstanding, most of
the results properly transcribed to the case of the real-valued
classical Klein-Gordon equation remain very similar in appearance to
the formulas that run from Eq.~(6a) through Eq.~(8f).

In particular, Eq.~(6a) needs essentially no modification; one simply
bears in mind that the operator $K$ is given by Eq.~(9b), and one re%
places the occurrences of $q$ and $\dot q$ by $\phi$ and $\dot\phi$.
The only remaining issue is one of a global reconciliation of dimen%
sion, which requires the determination of the parameter $\gm$ that ap%
pears Eq.~(6a) so as to accord with the conventions one intends to a%
dopt for the classical Klein-Gordon theory.  Now one \e{conventional}
choice of dimension for $\phi$ is the \e{same} as that of the electro%
magnetic vector potential $\vA$~[3, 1], which implies that
$\int|\del\phi|^2 d^3\vr$ has the dimension of energy.  A glance at
the classical conserved energy given by Eq.~(6c) reveals that $\gm$
must have the dimension of $c$, so we choose the value $c$ for $\gm$.
With that, Eq.~(6a) yields the mapping into the wave function and Ham%
iltonian operator of the \Schr\ equation that corresponds to the clas%
sical Klein-Gordon theory,
\re{
    \psi = (K^\qrs\phi + iK^{-\qrs}\dot\phi)/(2c^2\h)^\hfs,\quad
     H = \h K^\hfs,
}{9c}
where the operator $K$ is, of course, given by Eq.~(9b).  The \e{in%
verse} of this mapping from $\phi$ and $\dot\phi$ into the complex-%
valued \Schr\ wave function $\psi$ is easily calculated, or may be
transcribed from Eq.~(6e),
\re{
     \phi  =   c(\h/2)^\hfs K^{-\qrs}(\psi + \psi^\ast),\quad
  \dot\phi = -ic(\h/2)^\hfs K^\qrs(\psi - \psi^\ast).
}{9d}
Now let's take a closer look at the \Schr\ equation's Hamiltonian
operator,
\re{
    H = \h K^\hfs = \h(-c^2\del^2 + ((mc^2)/\h)^2)^\hf
}{9e}
In configuration space the quantum momentum operator $\qvp$ is well-%
known to be given by,
\re{
\qvp = -i\h\del,
}{9f}
so that,
\re{
    -\del^2 = |\qvp|^2/\h^2,
}{9g}
which, when substituted into the expression for $H$ in Eq.~(9e),
yields,
\re{
    H = (|c\qvp|^2 + m^2c^4)^\hf,
}{9h}
which is precisely the quantization of the standard relativistic ener%
gy of a free particle of mass $m$.  Thus we have the fascinating state
of affairs that the \e{classical} Klein-Gordon equation (i.e., with
\e{real-valued} $\phi$) is \e{linearly isomorphic} to the very \Schr\
equation with the correspondence-principle \e{mandated} square-root
Hamiltonian for the free particle of mass $m$ that Klein and Gordon
were in fact trying to sideline.  If Klein and Gordon had but been
aware of the Eq.~(6a) theorem with its \e{square-root character} of
the Hamiltonian matrix $H = \h K^\hfs$, the history of relativistic
quantum mechanics and its second quantization might have taken a dif%
ferent route, one in closer harmony with the correspondence princi%
ple.

Second quantization of the \Schr\ wave function $\psi$ for the classi%
cal Klein-Gordon theory can be transcribed from Eqs.~(8).  Here the
different dimension of $\psi$ that is imposed by its continuum charac%
ter results in its basic commutation relations coming out in terms of
a three-dimensional delta function instead of in terms of the Krone%
cker delta of Eq.~(8b).
\re{
   [\q\psi(\vr), \q\psi^\dg(\vr')] = \dl^{(3)}\!(\vr - \vr'),\qquad
   [\q\psi(\vr), \q\psi(\vr')] = 0,\qquad
   [\q\psi^\dg(\vr), \q\psi^\dg(\vr')] = 0.
}{9i}
This promotion of the \e{\Schr\ wave function} $\psi(\vr)$ to operator
field is the most straightforward and physically transparent route to
the \e{quantization} of the classical Klein-Gordon field $\phi(\vr)$,
which, of course, is explicitly given by Eq.~(9d) in terms of the
\Schr\ wave function and its complex conjugate.  The familiar physical
interpretation attached to the commutation relations of Eq.~(9i) is
that the operator field $\q\psi^\dg(\vr)$ creates a relativistic spin%
less particle of mass $m$ at location $\vr$, while the operator field
$\q\psi(\vr)$ destroys such a particle.  Such particle creation and
destruction operator fields are \e{non-Hermitian}.  However, from the
first equality of Eq.~(9d) we note that the quantized classical Klein%
-Gordon field $\q\phi(\vr)$ \e{itself} will, on the contrary, turn out
to be \e{Hermitian}, and will be ambiguously capable of \e{both} par%
ticle creation and annihilation.  In light of the second equality in
Eq.~(9d), the same comments apply to the quantization of the time der%
ivative of the classical Klein-Gordon field $d\q\phi(\vr)/dt$.  A tel%
ling characteristic of both of these \e{Hermitian} fields is that
\e{by themselves} they \e{only} obey the original \e{second-order}
real-valued \e{classical} Klein-Gordon equation.  Eqs.~(9c), (9d) and
(9i) thus mathematically depict the \e{complementarity} of the quan%
tized particle outlook (oriented toward non-Hermitian second-quantized
wave-functions that unambiguously either annihilate or create parti%
cles, and obey a first-order complex-valued \e{quantum} \Schr\ equa%
tion) to the classical wave outlook (oriented toward Hermitian fields
that by themselves \e{only} obey a real-valued second-order \e{classi%
cal} wave equation).

Finally, we wish to exhibit, in terms of these quantized \Schr\ wave
functions that create or destroy particles, the Hamiltonian operator
functional that oversees free relativistic spinless particles in the
second quantized world (we already met this operator in schematic form
in Eq.~(8d)),
\re{
    \qH[\qpsi, \qpsid] =
          \hf\tint[\qpsid(\vr)(-c^2\h^2\del^2 +m^2c^4)^\hf\qpsi(\vr) +
          \qpsi(\vr)(-c^2\h^2\del^2 +m^2c^4)^\hf\qpsid(\vr)]d^3\vr.
}{9j}

We now turn to the similar \Schr\ equation that corresponds to the
real-valued homogeneous linear source-free Maxwell equations.  The
differences to the \Schr-equation results for the classical Klein-Gor%
don equation are that the resulting relativistic particle is \e{mass%
less}, and that its wave function is a \e{vector} field which is
\e{strictly transverse}.

\sct{Free-photon quantum mechanics from the source-free Maxwell equations}
\noindent
In the source-free case, the Coulomb and Gauss laws tell us that both
the electric and magnetic fields are purely \e{transverse}, i.e.,
$\del\dt\vE = 0$ and $\del\dt\vB = 0$.  The results of the Maxwell law
and  Faraday's law in the source-free case are,
\re{
    \dot\vE = c\del\x\vB,\qquad \dot\vB = -c\del\x\vE.
}{10a}
This first-order equation system has the simple antisymmetrical char%
acter of Eq.~(7b), which readily produces a \Schr\ equation.  For ex%
ample, the extremely simple transverse-vector wave function \e{ansatz}
$\vPsi = \vE + i\vB$ will in consequence of Eq.~(10a) satisfy the
\Schr\ equation which has the Hamiltonian operator $\h c\,\bv{curl}$.
Unfortunately this operator has \e{odd parity}, and therefore is not a
physically appropriate Hamiltonian for electromagnetism.  The reason
that a Hamiltonian of odd parity has manifested itself here is that
the transverse vector fields on either side of each of the two equa%
tions of Eq.~(10a) are of \e{opposite} intrinsic parity: namely $\vE$
is a \e{polar} vector field, while $\vB$ is an \e{axial} vector field.
So it should be feasible to extract a physically appropriate \e{even-%
parity} \Schr-equation Hamiltonian operator from source-free electro%
magnetic theory by \e{first} recasting its linear homogeneous equa%
tions of motion such that they involve \e{only} transverse vector
fields \e{which all have the same intrinsic parity}.  We shall do this
here by mapping the transverse axial-vector magnetic field $\vB$ into
a transverse polar-vector field that is already well-known to electro%
magnetic theory, namely the vector potential in radiation gauge~[4].
Specifically, we define,
\re{
    \vA\eqdf (-\del^2)^{-1}(\del\x\vB),
}{10b}
where by $(-\del^2)^{-1}$ we mean the standard real-valued symmetric
\e{integral operator} with the Coulomb kernel.  Eq.~(10b) implies
that,
\re{
    \del\dt\vA = 0,
}{10c}
i.e., $\vA$ is a \e{transverse} vector field.  Furthermore, since
$\vB$ is \e{itself} a transverse vector field, Eq.~{10b} implies
that,
\re{
    \del\x\vA = \vB,
}{10d}
which is, of course, the basic property of a vector potential $\vA$.
We can further delineate the properties of $\vA$ in source-free elec%
tromagnetism by using its definition together with Faraday's law
(i.e., the second equality in Eq.~(10a)) to calculate its time deriva%
tive,
\re{
    \dot\vA = (-\del^2)^{-1}(\del\x\dot\vB) =
            -c(-\del^2)^{-1}(\del\x(\del\x\vE)) = -c\vE,
}{10e}
where the last equality holds when $\vE$ is \e{transverse}, which is,
of course the case for source-free electromagnetism.  So in that case,
\re{
    \vE = -\dot\vA/c.
}{10f}
Eqs.~(10d) and (10f) together imply that for source-free electromagne%
tism, we can obtain \e{both} of $\vB$ and $\vE$ from $\vA$, so we
\e{only} need to concern ourselves with calculating the polar trans%
verse vector field $\vA$.  Therefore we now substitute Eqs.~(10d) and
(10f) into the Maxwell law, which in the case of source-free electro%
magnetism is the first equality of Eq.~(10a), to obtain a linear homo%
geneous second-order equation which involves the \e{polar} transverse
vector field $\vA$ \e{alone},
\re{
    \ddot\vA - c^2\del^2\vA = 0.
}{10g}
This is, of course, the \e{classical wave equation}, and it bears a
\e{marked resemblance} to the classical Klein-Gordon equation of
Eq.~(9a).  The \e{only} differences are that in Eq.~(10g) the param%
eter $\om$ that appears in Eq.~(9a) \e{vanishes identically}, and,
of course, in Eq.~(10g) the transverse vector field $\vA$ \e{replaces}
the scalar field $\phi$ of Eq.~(9a).  Even the \e{dimension} of the
transverse vector field $\vA$ is the \e{same} as the dimension that we
\e{chose} for $\phi$ by adhering to a common convention~[3, 1].
Therefore, for the linear mapping, and its inverse, of the real-valued
transverse-vector fields $\vA$ and $\dot\vA$ into a complex-valued
transverse-vector \Schr-equation wave function $\vPsi$, we can simply
transcribe Eqs.~(9b), (9c) and (9d) for the real-valued classical
scalar Klein-Gordon theory, taking $\om$ (and $m$) to be zero identi%
cally, and replacing $\phi$, $\dot\phi$, and $\psi$ by, respectively,
$\vA$, $\dot\vA$, and $\vPsi$.  Thus our basic real, symmetric operator
is,
\re{
    K = -c^2\del^2,
}{10h}
which, to be sure, is not positive-definite in the broadest sense.
However, Fourier transformation methodology indicates that on a suffi%
ciently restricted function space, $-\del^2$ can indeed be regarded as
positive definite.  The operators we \e{actually require} in the fol%
lowing mapping formulas are $(-\del^2)^\hf$, $(-\del^2)^{-\qr}$ and
$(-\del^2)^\qr$, and they \e{themselves} have the tractable-looking
positive-definite Fourier representations $|\vk|$, $|\vk|^{-\hf}$ and
$|\vk|^\hf$ respectively.

Transcribing Eq.~(9c) as described above, the linear mapping of the
real-valued transverse-vector fields $\vA$ and $\dot\vA$ into the com%
plex-valued transverse-vector \Schr-equation wave function $\vPsi$,
together with the associated \Schr-equation Hamiltonian operator, is
given by,
\re{
    \vPsi = (K^\qrs\vA + iK^{-\qrs}\dot\vA)/(2c^2\h)^\hfs,\quad
     H = \h K^\hfs.
}{10i}
The \e{inverse} of this linear mapping from $\vA$ and $\dot\vA$ into
the complex-valued \Schr-equation wave function $\vPsi$ is,
\re{
     \vA  =   c(\h/2)^\hfs K^{-\qrs}(\vPsi + \vPsi^\ast),\quad
  \dot\vA = -ic(\h/2)^\hfs K^\qrs(\vPsi - \vPsi^\ast).
}{10j}

In light of Eq.~(10h) and the fact that in configuration representa%
tion $\qvp = -i\h\del$, we have from the second equality in Eq.~(10i)
that the \Schr-equation Hamiltonian operator can be written,
\re{
    H = \h K^\hfs = \h(-c^2\del^2)^\hf = (c^2|\qvp|^2)^\hf = c|\qvp|.
}{10k}
This Hamiltonian operator is clearly the quantized version of the rel%
ativistic energy of a \e{massless} free particle, which is appropriate
to the free photon, and it as well has even parity.

By using Eqs.~(10b) and (10e), the vector potential can be removed
from the \Schr-equation linear mapping of Eq.~(10i) in favor of the
$\vE$ and $\vB$ fields,
\re{
    \vPsi = (cK^{-\tqrs}(\del\x\vB) - iK^{-\qrs}\vE)/(2\h)^\hfs,\quad
     H = \h K^\hfs.
}{11a}
The mapping of $\vE$ and $\vB$ into $\vPsi$ given in Eq.~(11a) has the
inverse,
\re{
     \vB  =   c(\h/2)^\hfs K^{-\qrs}(\del\x(\vPsi + \vPsi^\ast)),\quad
     \vE =    i(\h/2)^\hfs K^\qrs(\vPsi - \vPsi^\ast).
}{11b}
We invite the reader to \e{verify} that the complex-valued linear map%
ping of the classical $\vE$ and $\vB$ fields into the wave function
$\vPsi$ which Eq.~(11a) specifies, along with its specified Hamilton%
ian operator $H = \h K^\hfs$ (where $K = -c^2\del^2$), \e{actually sa%
tisfies} the \Schr\ equation.  (Hint: use the source-free Maxwell and
Faraday laws of Eq.~(10a) and the transverse nature of the source-free
$\vE$ field.)  One should \e{also verify} that the quantum expectation
value of the Hamiltonian operator \e{agrees} with the \e{classical
energy} of the $\vE$ and $\vB$ field system, i.e., that,
\re{
    \tint\vPsi^\ast(\vr)\dt\lf(H\vPsi(\vr)\rt)d^3\vr =
    \hf\tint\lf(|\vE(\vr)|^2 + |\vB(\vr)|^2\rt)d^3\vr.
}{11c}

Upon their second quantization, Eqs.~(11a) and (11b) manifest the ex%
pected tantalizing \e{complementary} interplay of the potential for
photon creation and annihilation with the familiar, workaday trans%
verse electric and magnetic fields.

In addition to its zero mass parameter, the second special feature
of electromagnetic theory vis-\`a-vis classical Klein-Gordon theory
is, of course, the free photon's \e{always transverse} polarization
(spin) states.  This signature free-photon characteristic does not
cause much in the way of complications, but there is one formula
concerning second quantization which it \e{notationally} impacts,
albeit \e{no substantive physical effect is involved}.  The canoni%
cal commutation rule for second quantization of the free photon's
transverse vector wave function might naively be expected to read,
\re{
 [(\qvPsi(\vr))_i, (\qvPsi^\dg(\vr'))_j] = \dl_{ij}\dl^{(3)}\!(\vr - \vr'),
}{12a}
but this is \e{not} mathematically consistent with the transverse
character of the second-quantized photon wave-functions, i.e., it is
mathematically inconsistent with the fact that $\del\dt\qvPsi = 0$.
The nature of the right-hand of Eq.~(12a) is one of completeness, but
the transverse wave function creation and annihilation operators are
\e{incomplete} in that they do \e{not} pertain to vector fields which
are the gradients of scalar fields, i.e., they do \e{not} pertain to
vector fields \e{which fail to be transverse}.  Now the $ij$ compon%
ents of the \e{projection operator} onto the subspace of such purely
gradient vector fields is given by,
\re{
    P_{ij} = -\pa_i (-\del^2)^{-1}\pa_j.
}{12b}
We note that $P_{ij}$ is Hermitian, and that its contraction with it%
self yields itself, which are the two essential properties of the $ij$
components of projection operators.  Of course its contraction with
the components of any transverse vector field vanishes.  Thus
$(\dl_{ij} - P_{ij})$ are the $ij$ components of the \e{projection op%
erator onto the subspace of transverse vector fields}, and therefore,
\re{
 [(\qvPsi(\vr))_i, (\qvPsi^\dg(\vr'))_j] = \la\vr|(\dl_{ij} - P_{ij})|\vr'\ra
  = (2\pi)^{-3}\int e^{i\vk\dt(\vr - \vr')}
     \lf(\dl_{ij}  - \vk_i\vk_j|\vk|^{-2}\rt)d^3\vk.
}{12c}
Notwithstanding these fancy maneuvers with projection operators, the
\e{only} issue which is involved here is the simple fact that free-%
photon creation and annihilation operators (and as well free photon
wave functions in the first quantized regime) are \e{purely trans%
verse}, and therefore \e{any expression involving these operators},
e.g., the expression which describes their canonical commutation
relation, must, of course, \e{correctly reflect this fact}.  There
is obviously \e{no physics implication} which flows from this re%
quirement of \e{mere notational correctness}.

\sct{Conclusion}
\noindent
It is a remarkable fact that any classical system whose equation of
motion is linear, homogeneous, \e{purely} oscillatory and conservative
is effectively {\em already first-quantized}: once its Eq.~(1a) coup%
ling-strength matrix $K$ has been similarity-transformed to a symmet%
ric, positive-definite presentation, Eq.~(6a) invertibly linearly maps
that equation of motion into explicit time-dependent Schr\"{o}dinger-%
equation form with Hamiltonian matrix $\,\h K^\hfs$.  Thus we see that
Michael Faraday and James Clerk Maxwell were actually the first to ef%
fectively elucidate a quantized particle, namely the very important
and not exactly simple ultra-relativistic massless transverse-vec%
tor free photon.

Any \e{complex-valued} solution wave function of a time-dependent
Schr\"{o}dinger-equation has the familiar characteristic expansion in
terms of the complete set of mutually orthogonal eigenfunctions of
that equation's Hamiltonian operator.  The one-to-one linear mapping
of any purely oscillatory linear classical system that is homogeneous
and conservative into a Schr\"{o}dinger equation thus implies a char%
acteristic \e{two-component} eigenfunction expansion of such a classi%
cal system's solutions.  For the case of certain wave equations that
fall into the class of Eq.~(1a), precisely such a solution expansion
has been described in detail by Leung, Tong and Young~[5].

The natural \e{correspondence-principle} version of the relativistic
free-particle Schr\"{o}dinger equation was \e{iterated} by Klein,
Gordon and Schr\"{o}dinger \e{for no physically motivated reason}, but
merely in an effort to \e{rid} it of its \e{calculationally unpalata%
ble} square-root Hamiltonian operator~[6, 1, 7].  If this \e{iterated}
equation is \e{still} regarded as a \e{complex-valued quantum-mechani%
cal entity}, a large class of \e{completely extraneous, highly unphys%
ical unbounded-below negative-energy solutions} are \e{injected} by
that iteration.  These also destroy its probability interpretation,
and the fact that it depends on only the \e{square} of a Hamiltonian
cuts it adrift from the Heisenberg picture and Ehrenfest theorem.
However, if this iterated equation is regarded as the description of a
\e{classical, real-valued} field, it thereupon becomes strongly analo%
gous to the \e{classical wave equation}, and has an eminently sensible
\e{nonnegative} conserved energy~[3, 1].  This \e{classical} Klein-%
Gordon equation is \e{as well} one of those classical equation systems 
which is linearly equivalent to a Schr\"{o}dinger equation: it quite
marvelously \e{chooses} to be equivalent to \e{precisely} the
Schr\"{o}dinger equation with the natural correspondence-principle
\e{square-root Hamiltonian operator} which Klein, Gordon and
Schr\"{o}dinger \e{had tried to sideline by concocting it}.

It is a pity that Klein, Gordon and Schr\"{o}dinger had no idea of the
theorem presented by this paper, and thus were not equipped to unearth
this astonishing fact themselves.  If they had but grasped the full
consequences of the real-valued classical Klein-Gordon equation, they
might well have abandoned their physically unmotivated rejection of
the \e{correspondence-principle mandated} relativistic free-particle
square-root Hamiltonian operator $(|c\qvp|^2 + m^2c^4)^\hf$~[7, 1].

\vskip 1.75\baselineskip\noindent{\frtbf References}
\vskip 0.25\baselineskip

{\parindent = 15pt
\sk\item{[1]}
S. K. Kauffmann,
arXiv:1012.5120 [physics.gen-ph]
(2010).
\sk\item{[2]}
S. K. Kauffmann,
arXiv:1011.6578 [physics.gen-ph]
(2010).
\sk\item{[3]}
S. S. Schweber,
\e{An Introduction to Relativistic Quantum Field Theory}
(Harper \& Row, New York, 1961).
\sk\item{[4]}
J. D. Bjorken and S. D. Drell,
\e{Relativistic Quantum Fields}
(McGraw-Hill, New York, 1965).
\sk\item{[5]}
P. T. Leung, S. S. Tong and K. Young,
J.\ Phys.\ A: Math.\ Gen.\ \textbf{30},
2139 (1997).
\sk\item{[6]}
J. D. Bjorken and S. D. Drell,
\e{Relativistic Quantum Mechanics}
(McGraw-Hill, New York, 1964).
\sk\item{[7]}
S. K. Kauffmann,
arXiv:1009.3584 [physics.gen-ph]
(2010).
}
\bye